


\message
{JNL.TEX version 0.92 as of 6/9/87.  Report bugs and problems to Doug Eardley.}

\catcode`@=11
\expandafter\ifx\csname inp@t\endcsname\relax\let\inp@t=\input
\def\input#1 {\expandafter\ifx\csname #1IsLoaded\endcsname\relax
\inp@t#1%
\expandafter\def\csname #1IsLoaded\endcsname{(#1 was previously loaded)}
\else\message{\csname #1IsLoaded\endcsname}\fi}\fi
\catcode`@=12



\font\twelverm=cmr10 scaled 1200    \font\twelvei=cmmi10 scaled 1200
\font\twelvesy=cmsy10 scaled 1200   \font\twelveex=cmex10 scaled 1200
\font\twelvebf=cmbx10 scaled 1200   \font\twelvesl=cmsl10 scaled 1200
\font\twelvett=cmtt10 scaled 1200   \font\twelveit=cmti10 scaled 1200
\font\twelvesc=cmcsc10 scaled 1200  \font\twelvesf=amssmc10 scaled 1200
\skewchar\twelvei='177   \skewchar\twelvesy='60


\def\twelvepoint{\normalbaselineskip=12.4pt plus 0.1pt minus 0.1pt
  \abovedisplayskip 12.4pt plus 3pt minus 9pt
  \belowdisplayskip 12.4pt plus 3pt minus 9pt
  \abovedisplayshortskip 0pt plus 3pt
  \belowdisplayshortskip 7.2pt plus 3pt minus 4pt
  \smallskipamount=3.6pt plus1.2pt minus1.2pt
  \medskipamount=7.2pt plus2.4pt minus2.4pt
  \bigskipamount=14.4pt plus4.8pt minus4.8pt
  \def\rm{\fam0\twelverm}          \def\it{\fam\itfam\twelveit}%
  \def\sl{\fam\slfam\twelvesl}     \def\bf{\fam\bffam\twelvebf}%
  \def\mit{\fam 1}                 \def\cal{\fam 2}%
  \def\sc{\twelvesc}		   \def\tt{\twelvett}
  \def\sf{\twelvesf}
  \textfont0=\twelverm   \scriptfont0=\tenrm   \scriptscriptfont0=\sevenrm
  \textfont1=\twelvei    \scriptfont1=\teni    \scriptscriptfont1=\seveni
  \textfont2=\twelvesy   \scriptfont2=\tensy   \scriptscriptfont2=\sevensy
  \textfont3=\twelveex   \scriptfont3=\twelveex  \scriptscriptfont3=\twelveex
  \textfont\itfam=\twelveit
  \textfont\slfam=\twelvesl
  \textfont\bffam=\twelvebf \scriptfont\bffam=\tenbf
  \scriptscriptfont\bffam=\sevenbf
  \normalbaselines\rm}



\def\beginlinemode{\endmode
  \begingroup\parskip=0pt \obeylines\def\\{\par}\def\endmode{\par\endgroup}}
\def\beginparmode{\endmode
  \begingroup \def\endmode{\par\endgroup}}
\let\endmode=\par
{\obeylines\gdef\
{}}
\def\singlespace{\baselineskip=\normalbaselineskip}

\def\oneandahalfspace{\baselineskip=\normalbaselineskip
  \multiply\baselineskip by 3 \divide\baselineskip by 2}
\def\doublespace{\baselineskip=\normalbaselineskip \multiply\baselineskip by 2}

\newcount\firstpageno
\firstpageno=2
\footline={\ifnum\pageno<\firstpageno{\hfil}\else{\hfil\twelverm\folio\hfil}\fi}
\def\toppageno{\global\footline={\hfil}\global\headline
  ={\ifnum\pageno<\firstpageno{\hfil}\else{\hfil\twelverm\folio\hfil}\fi}}
\let\rawfootnote=\footnote		
\def\footnote#1#2{{\rm\singlespace\parindent=0pt\parskip=0pt
  \rawfootnote{#1}{#2\hfill\vrule height 0pt depth 6pt width 0pt}}}
\def\raggedcenter{\leftskip=4em plus 12em \rightskip=\leftskip
  \parindent=0pt \parfillskip=0pt \spaceskip=.3333em \xspaceskip=.5em
  \pretolerance=9999 \tolerance=9999
  \hyphenpenalty=9999 \exhyphenpenalty=9999 }
\def\dateline{\rightline{\ifcase\month\or
  January\or February\or March\or April\or May\or June\or
  July\or August\or September\or October\or November\or December\fi
  \space\number\year}}
\def\received{\vskip 3pt plus 0.2fill
 \centerline{\sl (Received\space\ifcase\month\or
  January\or February\or March\or April\or May\or June\or
  July\or August\or September\or October\or November\or December\fi
  \qquad, \number\year)}}


\hsize=6.5truein
\hoffset=0truein
\vsize=8.9truein
\voffset=0truein
\parskip=\medskipamount
\def\\{\cr}
\twelvepoint		
\doublespace		
\overfullrule=0pt	




\def\title			
  {\null\vskip 3pt plus 0.2fill
   \beginlinemode \doublespace \raggedcenter \bf}

\def\author			
  {\vskip 3pt plus 0.2fill \beginlinemode
   \singlespace \raggedcenter\sc}

\def\affil			
  {\vskip 3pt plus 0.1fill \beginlinemode
   \oneandahalfspace \raggedcenter \sl}

\def\abstract			
  {\vskip 3pt plus 0.3fill \beginparmode
   \oneandahalfspace ABSTRACT: }

\def\endtitlepage		
  {\endpage			
   \body}
\let\endtopmatter=\endtitlepage

\def\body			
  {\beginparmode}		

\def\head#1{			
  \goodbreak\vskip 0.5truein	
  {\immediate\write16{#1}
   \raggedcenter \uppercase{#1}\par}
   \nobreak\vskip 0.25truein\nobreak}

\def\beginitems{
\par\medskip\bgroup\def\i##1 {\item{##1}}\def\ii##1 {\itemitem{##1}}
\leftskip=36pt\parskip=0pt}
\def\enditems{\par\egroup}

\def\beneathrel#1\under#2{\mathrel{\mathop{#2}\limits_{#1}}}

\def\refto#1{$^{#1}$}		

\def\references			
  {\head{References}		
   \beginparmode
   \frenchspacing \parindent=0pt \leftskip=1truecm
   \parskip=8pt plus 3pt \everypar{\hangindent=\parindent}}

\def\referencesnohead   	
  {                     	
   \beginparmode
   \frenchspacing \parindent=0pt \leftskip=1truecm
   \parskip=8pt plus 3pt \everypar{\hangindent=\parindent}}

\gdef\refis#1{\item{#1.\ }}			

\gdef\journal#1, #2, #3, 1#4#5#6{		
    {\sl #1~}{\bf #2}, #3 (1#4#5#6)}		

\def\pr{\journal Phys. Rev., }

\def\prb{\journal Phys. Rev. B, }

\def\prl{\journal Phys. Rev. Lett., }

\def\np{\journal Nucl. Phys., }

\def\pl{\journal Phys. Lett., }

\def\endreferences{\body}

\def\figurecaptions		
  {\endpage
   \beginparmode
   \head{Figure Captions}
}

\def\endpage			
  {\vfill\eject}

\def\endpaper			
  {\endmode\vfill\supereject}


\def\heading				
  {\vskip 0.5truein plus 0.1truein	
   \beginparmode \def\\{\par} \parskip=0pt \singlespace \raggedcenter}

\def\subheading				
  {\vskip 0.25truein plus 0.1truein	
   \beginlinemode \singlespace \parskip=0pt \def\\{\par}\raggedcenter}

\def\tag#1$${\eqno(#1)$$}

\def\align#1$${\eqalign{#1}$$}

\def\aligntag#1$${\gdef\tag##1\\{&(##1)\cr}\eqalignno{#1\\}$$
  \gdef\tag##1$${\eqno(##1)$$}}

\def\endaligntag{}

\def\overset #1\to#2{{\mathop{#2}\limits^{#1}}}
\def\underset#1\to#2{{\let\next=#1\mathpalette\undersetpalette#2}}
\def\undersetpalette#1#2{\vtop{\baselineskip0pt
\ialign{$\mathsurround=0pt #1\hfil##\hfil$\crcr#2\crcr\next\crcr}}}


\def\ref#1{Ref.~#1}			
\def\Ref#1{Ref.~#1}			
\def\[#1]{[\cite{#1}]}
\def\cite#1{{#1}}
\def\(#1){(\call{#1})}
\def\call#1{{#1}}
\def\taghead#1{}
\def\frac#1#2{{#1 \over #2}}

\def\12{{1\over2}}

\def\ie{{\it i.e.,\ }}

\def\etc{{\it etc.\ }}

\def\sla{\raise.15ex\hbox{$/$}\kern-.57em}
\def\leaderfill{\leaders\hbox to 1em{\hss.\hss}\hfill}
\def\twiddle{\lower.9ex\rlap{$\kern-.1em\scriptstyle\sim$}}
\def\bigtwiddle{\lower1.ex\rlap{$\sim$}}
\def\gtwid{\mathrel{\raise.3ex\hbox{$>$\kern-.75em\lower1ex\hbox{$\sim$}}}}
\def\ltwid{\mathrel{\raise.3ex\hbox{$<$\kern-.75em\lower1ex\hbox{$\sim$}}}}
\def\square{\kern1pt\vbox{\hrule height 1.2pt\hbox{\vrule width 1.2pt\hskip 3pt
   \vbox{\vskip 6pt}\hskip 3pt\vrule width 0.6pt}\hrule height 0.6pt}\kern1pt}
\def\tdot#1{\mathord{\mathop{#1}\limits^{\kern2pt\ldots}}}

\def\pmb#1{\setbox0=\hbox{#1}%
  \kern-.025em\copy0\kern-\wd0
  \kern  .05em\copy0\kern-\wd0
  \kern-.025em\raise.0433em\box0 }

\catcode`@=11
\newcount\r@fcount \r@fcount=0
\newcount\r@fcurr
\immediate\newwrite\reffile
\newif\ifr@ffile\r@ffilefalse
\def\w@rnwrite#1{\ifr@ffile\immediate\write\reffile{#1}\fi\message{#1}}

\def\writer@f#1>>{}
\def\referencefile{
  \r@ffiletrue\immediate\openout\reffile=\jobname.ref%
  \def\writer@f##1>>{\ifr@ffile\immediate\write\reffile%
    {\noexpand\refis{##1} = \csname r@fnum##1\endcsname = %
     \expandafter\expandafter\expandafter\strip@t\expandafter%
     \meaning\csname r@ftext\csname r@fnum##1\endcsname\endcsname}\fi}%
  \def\strip@t##1>>{}}

\def\citeall#1{\xdef#1##1{#1{\noexpand\cite{##1}}}}
\def\cite#1{\each@rg\citer@nge{#1}}	

\def\each@rg#1#2{{\let\thecsname=#1\expandafter\first@rg#2,\end,}}
\def\first@rg#1,{\thecsname{#1}\apply@rg}	
\def\apply@rg#1,{\ifx\end#1\let\next=\relax
\else,\thecsname{#1}\let\next=\apply@rg\fi\next}

\def\citer@nge#1{\citedor@nge#1-\end-}	
\def\citer@ngeat#1\end-{#1}
\def\citedor@nge#1-#2-{\ifx\end#2\r@featspace#1 
  \else\citel@@p{#1}{#2}\citer@ngeat\fi}	
\def\citel@@p#1#2{\ifnum#1>#2{\errmessage{Reference range #1-#2\space is bad.}%
    \errhelp{If you cite a series of references by the notation M-N, then M and
    N must be integers, and N must be greater than or equal to M.}}\else%
 {\count0=#1\count1=#2\advance\count1
by1\relax\expandafter\r@fcite\the\count0,%
  \loop\advance\count0 by1\relax
    \ifnum\count0<\count1,\expandafter\r@fcite\the\count0,%
  \repeat}\fi}

\def\r@featspace#1#2 {\r@fcite#1#2,}	
\def\r@fcite#1,{\ifuncit@d{#1}
    \newr@f{#1}%
    \expandafter\gdef\csname r@ftext\number\r@fcount\endcsname%
                     {\message{Reference #1 to be supplied.}%
                      \writer@f#1>>#1 to be supplied.\par}%
 \fi%
 \csname r@fnum#1\endcsname}
\def\ifuncit@d#1{\expandafter\ifx\csname r@fnum#1\endcsname\relax}%
\def\newr@f#1{\global\advance\r@fcount by1%
    \expandafter\xdef\csname r@fnum#1\endcsname{\number\r@fcount}}

\let\r@fis=\refis			
\def\refis#1#2#3\par{\ifuncit@d{#1}
   \newr@f{#1}%
   \w@rnwrite{Reference #1=\number\r@fcount\space is not cited up to now.}\fi%
  \expandafter\gdef\csname r@ftext\csname r@fnum#1\endcsname\endcsname%
  {\writer@f#1>>#2#3\par}}

\def\ignoreuncited{
   \def\refis##1##2##3\par{\ifuncit@d{##1}%
     \else\expandafter\gdef\csname r@ftext\csname
r@fnum##1\endcsname\endcsname%
     {\writer@f##1>>##2##3\par}\fi}}

\def\r@ferr{\endreferences\errmessage{I was expecting to see
\noexpand\endreferences before now;  I have inserted it here.}}
\let\r@ferences=\references
\def\references{\r@ferences\def\endmode{\r@ferr\par\endgroup}}

\let\endr@ferences=\endreferences
\def\endreferences{\r@fcurr=0
  {\loop\ifnum\r@fcurr<\r@fcount
    \advance\r@fcurr by 1\relax\expandafter\r@fis\expandafter{\number\r@fcurr}%
    \csname r@ftext\number\r@fcurr\endcsname%
  \repeat}\gdef\r@ferr{}\endr@ferences}


\let\r@fend=\endpaper\gdef\endpaper{\ifr@ffile
\immediate\write16{Cross References written on []\jobname.REF.}\fi\r@fend}

\catcode`@=12

\citeall\refto		
\citeall\ref		%
\citeall\Ref		%

\catcode`@=11
\newcount\tagnumber\tagnumber=0

\immediate\newwrite\eqnfile
\newif\if@qnfile\@qnfilefalse
\def\write@qn#1{}
\def\writenew@qn#1{}
\def\w@rnwrite#1{\write@qn{#1}\message{#1}}
\def\@rrwrite#1{\write@qn{#1}\errmessage{#1}}

\def\taghead#1{\gdef\t@ghead{#1}\global\tagnumber=0}
\def\t@ghead{}

\expandafter\def\csname @qnnum-3\endcsname
  {{\t@ghead\advance\tagnumber by -3\relax\number\tagnumber}}
\expandafter\def\csname @qnnum-2\endcsname
  {{\t@ghead\advance\tagnumber by -2\relax\number\tagnumber}}
\expandafter\def\csname @qnnum-1\endcsname
  {{\t@ghead\advance\tagnumber by -1\relax\number\tagnumber}}
\expandafter\def\csname @qnnum0\endcsname
  {\t@ghead\number\tagnumber}
\expandafter\def\csname @qnnum+1\endcsname
  {{\t@ghead\advance\tagnumber by 1\relax\number\tagnumber}}
\expandafter\def\csname @qnnum+2\endcsname
  {{\t@ghead\advance\tagnumber by 2\relax\number\tagnumber}}
\expandafter\def\csname @qnnum+3\endcsname
  {{\t@ghead\advance\tagnumber by 3\relax\number\tagnumber}}

\def\equationfile{%
  \@qnfiletrue\immediate\openout\eqnfile=\jobname.eqn%
  \def\write@qn##1{\if@qnfile\immediate\write\eqnfile{##1}\fi}
  \def\writenew@qn##1{\if@qnfile\immediate\write\eqnfile
    {\noexpand\tag{##1} = (\t@ghead\number\tagnumber)}\fi}
}

\def\callall#1{\xdef#1##1{#1{\noexpand\call{##1}}}}
\def\call#1{\each@rg\callr@nge{#1}}

\def\each@rg#1#2{{\let\thecsname=#1\expandafter\first@rg#2,\end,}}
\def\first@rg#1,{\thecsname{#1}\apply@rg}
\def\apply@rg#1,{\ifx\end#1\let\next=\relax%
\else,\thecsname{#1}\let\next=\apply@rg\fi\next}

\def\callr@nge#1{\calldor@nge#1-\end-}
\def\callr@ngeat#1\end-{#1}
\def\calldor@nge#1-#2-{\ifx\end#2\@qneatspace#1 %
  \else\calll@@p{#1}{#2}\callr@ngeat\fi}
\def\calll@@p#1#2{\ifnum#1>#2{\@rrwrite{Equation range #1-#2\space is bad.}
\errhelp{If you call a series of equations by the notation M-N, then M and
N must be integers, and N must be greater than or equal to M.}}\else%
 {\count0=#1\count1=#2\advance\count1
by1\relax\expandafter\@qncall\the\count0,%
  \loop\advance\count0 by1\relax%
    \ifnum\count0<\count1,\expandafter\@qncall\the\count0,%
  \repeat}\fi}

\def\@qneatspace#1#2 {\@qncall#1#2,}
\def\@qncall#1,{\ifunc@lled{#1}{\def\next{#1}\ifx\next\empty\else
  \w@rnwrite{Equation number \noexpand\(>>#1<<) has not been defined yet.}
  >>#1<<\fi}\else\csname @qnnum#1\endcsname\fi}

\let\eqnono=\eqno
\def\eqno(#1){\tag#1}
\def\tag#1$${\eqnono(\displayt@g#1 )$$}

\def\aligntag#1\endaligntag
  $${\gdef\tag##1\\{&(##1 )\cr}\eqalignno{#1\\}$$
  \gdef\tag##1$${\eqnono(\displayt@g##1 )$$}}

\def\eqalignno#1{\displ@y \tabskip\centering
  \halign to\displaywidth{\hfil$\displaystyle{##}$\tabskip\z@skip
    &$\displaystyle{{}##}$\hfil\tabskip\centering
    &\llap{$\displayt@gpar##$}\tabskip\z@skip\crcr
    #1\crcr}}

\def\displayt@gpar(#1){(\displayt@g#1 )}

\def\displayt@g#1 {\rm\ifunc@lled{#1}\global\advance\tagnumber by1
        {\def\next{#1}\ifx\next\empty\else\expandafter
        \xdef\csname @qnnum#1\endcsname{\t@ghead\number\tagnumber}\fi}%
  \writenew@qn{#1}\t@ghead\number\tagnumber\else
        {\edef\next{\t@ghead\number\tagnumber}%
        \expandafter\ifx\csname @qnnum#1\endcsname\next\else
        \w@rnwrite{Equation \noexpand\tag{#1} is a duplicate number.}\fi}%
  \csname @qnnum#1\endcsname\fi}

\def\ifunc@lled#1{\expandafter\ifx\csname @qnnum#1\endcsname\relax}

\let\@qnend=\end\gdef\end{\if@qnfile
\immediate\write16{Equation numbers written on []\jobname.EQN.}\fi\@qnend}

\catcode`@=12


\def\ie{{\it i.e.,\ }}

\def\etc{{\it etc.\ }}

\def\>{\rangle}
\def\<{\langle}
\def\o{\over}

\def\ch{{\rm ch}}
\def\t{\tilde}
\def\prop{\propto}

\def\slD{\raise.15ex\hbox{$/$}\kern-.57em\hbox{$D$}}
\def\dsl{\raise.15ex\hbox{$/$}\kern-.57em\hbox{$\Delta$}}
\def\slp{{\raise.15ex\hbox{$/$}\kern-.57em\hbox{$\partial$}}}
\def\nsl{\raise.15ex\hbox{$/$}\kern-.57em\hbox{$\nabla$}}
\def\sla{\raise.15ex\hbox{$/$}\kern-.57em\hbox{$\rightarrow$}}
\def\slla{\raise.15ex\hbox{$/$}\kern-.57em\hbox{$\lambda$}}
\def\slb{\raise.15ex\hbox{$/$}\kern-.57em\hbox{$b$}}
\def\lnp{\raise.15ex\hbox{$/$}\kern-.57em\hbox{$p$}}
\def\lnk{\raise.15ex\hbox{$/$}\kern-.57em\hbox{$k$}}
\def\lnK{\raise.15ex\hbox{$/$}\kern-.57em\hbox{$K$}}
\def\lnq{\raise.15ex\hbox{$/$}\kern-.57em\hbox{$q$}}

\def\a{\alpha}
\def\be{{\beta}}

\def\de{{\delta}}

\def\th{{\theta}}

\def\la{\lambda}
\def\si{{\sigma}}

\def\cA{{\cal A}}

\def\part{\partial}

\def\dag{\dagger}

\def\abs{
         \vskip 3pt plus 0.3fill\beginparmode
         \doublespace ABSTRACT:\ }

\singlespace
\def\ch{\hbox{ch}}
\def\Ch{\hbox{Ch}}

\title Chiral Operator Product Algebra
and Edge Excitations of a Fractional Quantum Hall Droplet

\author Xiao-Gang Wen

\affil
Department of Physics, MIT
77 Massachusetts Avenue
Cambridge, MA 02139, U.S.A.

\author Yong-Shi Wu

\affil
Department of Physics, University of Utah
Salt Lake City, UT 84112, U.S.A.

\author Yasuhiro Hatsugai

\affil
Institute for Solid State Physics
University of Tokyo
7-22-1 Roppongi, Minato-ku, Tokyo 106, Japan

\abs{
In this paper we study the spectrum of low-energy
edge excitations of a fractional quantum Hall (FQH)
droplet. We show how to generate, by conformal
field theory (CFT) techniques, the many-electron wave
functions for the edge states. And we propose to
classify the spectrum of the edge states
by the same chiral operator product algebra (OPA)
that appears in the CFT description of the ground
state in the bulk. This bulk-edge correspondence is
suggested particularly for FQH systems that support
quasiparticle obeying non-abelian braid statistics,
including the $\nu=5/2$ Haldane-Rezayi state. Numerical
diagonalization to count the low-lying edge states
has been done for several non-abelian FQH systems,
showing good agreement in all cases
with the chiral OPA predictions.
The specific heat of the edge excitations in those
non-abelian states is also calculated.
}
\endtopmatter

\head{1. Introduction}

At low energies the dynamical degrees of freedom of
an incompressible fractional quantum Hall (FQH)
fluid live on its boundary.\refto{Halp} They give
rise to gapless edge excitations which play important
role in many phenomena in FQH
systems.\refto{Been,MacD,Wen0,Hald,Stone,WenRev}
In particular, it has been proposed\refto{prop}
that the study of edge states may be used to
characterize the topological order\refto{Wen1}
of an FQH state. (By topological order,
we mean  ``universality class''
of properties of an FQH state that reflect certain
internal structure of the state and are robust against weak
disorders and weak perturbations of electron interactions.
Up to now such properties are known to include the
quantum numbers (including statistics) of the
quasiparticle excitations, the ground state degeneracy
on a torus, as well as the spectrum of edge states.)

It has been shown\refto{Wen1,WZ} that the edge states
of abelian FQH states, i.e. those which support
quasiparticles obeying abelian braid statistics (such
as usual hierarchical FQH states), can be classified
by multiple U(1) current algebras which are characterized
by a symmetric $K$-matrix with integer elements. This $U(1)$
current algebra description is known to be closely
related to chiral bosons in $U(1)$ Gaussian conformal field
theory (CFT) on the edge\refto{Wen0,Stone,Naples,Cern}.
It is remarkable that the edge CFT (in 1+1 dimensional
Minkowskian space-time) turns out to be the
same as the CFT in the bulk (but in 2-dimensional
Euclidean plane) whose correlations give
the many-body FQH wave functions for the ground state or
quasiparticle excited states.\refto{MR,Fubini,Naples}
On the other hand, there may exist FQH states
that support quasiparticle excitations
obeying non-abelian braid statistics,\refto{MR,Wen3,WB} among
which the $\nu=5/2$ Haldane-Rezayi state is an
outstanding candidate. How to characterize the
spectrum of the edge states of such non-abelian FQH
states remains an open question. Some work in this
direction has been done in \ref{Wp}, which suggests
that, for the non-abelian Pfaffian state
proposed in \Ref{MR}, the edge excitations
are again described by the same
CFT that generates the bulk wave functions.

Recently two of us have shown\refto{WW} that the many-body
wave functions of several (presumably non-abelian) FQH
ground states, such as d-wave paired states
for spinless or spin-1/2 electrons, admit a CFT
description. Namely these FQH wave functions satisfy
conformal Ward identities\refto{BPZ,book}
and, therefore, can be identified
with correlations of a primary field in appropriate CFT.
We have explicitly constructed the closed algebra that is
generated by the electron operator(s), as primary field(s),
through operator product expansions (OPE). This algebra,
called by us the center algebra, forms a subalgebra of
the complete chiral operator product algebra (OPA) of the CFT.
We have been able to extend it to include some disorder
operators corresponding to quasiparticles, and
thus able to determine the quantum numbers of some
of quasiparticle excitations in these FQH systems.

In the same paper it was proposed that the topological order
of these FQH states should be characterized by the chiral
OPA that appears in their CFT description.
It is thus natural to see whether the gapless excitations
of these FQH systems on the edge can be described by
the same chiral OPA or not. Two questions immediately
come to our mind in this regard. First, can we generate
the many-body wave functions for the edge excitations
as correlations of the same CFT, the way similar
to generating those for the bulk ground
state and quasiparticle excitations?
Second, can we classify the spectrum of edge states by
(representations of) the same chiral OPA?
This paper reports our research on these two questions.

In Sec. 2 we start with a study of how to generate many-body
wave functions that describe edge excitations in the Laughlin
$1/m$-state by CFT techniques of inserting appropriate
screening operators at spatial infinity.
We show that the edge states are organized by representations
of the same chiral $U(1)$ current algebra that appears
in the CFT description of bulk states,
with the angular momentum of the edge state to be
identified with the descendant level in the representation.
Then in Sec. 3 we generalize these CFT techniques
to several d-wave paired FQH states,
especially to the Haldane-Rezayi
spin-singlet state which is studied in great detail.
In particular, the edge-state
spectrum is proposed to be classified by representations
of the same chiral OPA appearing in the CFT description for
the corresponding bulk states.
Thus the number of edge states with a given
angular momentum can be predicted from the formulas known
in CFT for characters of various chiral OPA.
In Sec. 4 we present results of numerical tests of these
predictions. Numerical diagonalization is done to count
the low-lying edge states for the d-wave paired states
for either spinless or spin-1/2 electrons. The results
show good agreement in all cases with the predictions
from the chiral OPA classification. Also the specific
heat of edge excitations in these FQH states is calculated,
with an asymptotic analysis of the number of edge states
with very large angular momenta.

We would like to stress that to relate the bulk CFT
that generate the bulk wave functions to the edge
CFT that describes the edge
spectrum, we need to introduce a concept of
minimal CFT for the bulk CFT. This is because
once a FQH wave function
can be written as a correlation function in a certain CFT,
then the wave function can also be expressed as a
correlation in many other CFT which contain
the original CFT. The minimal (bulk) CFT for a
FQH state not only reproduces the wave function,
it is also contained by any other CFT that reproduces the
wave function. The bulk CFT constructed in \Ref{WW}
is automatically the minimal one, because,
by construction, the CFT obtained contains the minimal
set of primary fields that are needed to construct
the wave function. Furthermore the energy-momentum
tensor of the minimal CFT can be expressed in terms of the
operators that generate the wave function. As we will
see in this paper, it is the chiral OPA in this minimal
CFT that describes the edge excitations
of the corresponding FQH state.

\head{2. Edge excitations and current algebra}

In this section we are going to study edge excitations
of the Laughlin state using a current algebra approach.

We note that the $\nu=1/m$ Laughlin state
(in the symmetric gauge)
$$\prod_{i<j} (z_i-z_j)^m e^{-{1\o 4}\sum_i|z_i|^2}
\equiv \Phi_m(z_i) e^{-{1\o 4}\sum_i|z_i|^2}
\eqno(1)$$
is a zero-energy state of a Hamiltonian
with the following two-body interaction\refto{TK}
$$V(z_1-z_2)=-\sum_{n=1}^{(m-1)/2} C_n \part_{z_1^*}^{2n-1}
\de (z_1-z_2) \part_{z_1}^{2n-1}
\eqno(2)$$
with $C_n>0$. (The Haldane's pseudo-potentials\refto{psd}
for the above
interaction are given by $V_{2l+1}>0$ for $2l+1<m$ and
 $V_{2l+1}=0$ for $2l+1\geq m$.)
There are also
many other zero-energy states for this Hamiltonian.
In fact any antisymmetric wave function of form
$$\t \Phi_m(z_i) e^{-{1\o 4}\sum_i|z_i|^2}
\eqno(3)$$
has zero energy, if and only if $\t \Phi_m$ does not contain
zeros of order less than $m$, \ie
$$\t\Phi_m(z_i)=O((z_1-z_2)^{m})
\eqno(4)$$
as $z_1\to z_2$.
The wave function \(1) describes a circular droplet of the
incompressible FQH fluid in its ground state.
Other zero-energy states correspond to edge-deformed
droplets, i.e. excited states of the droplet
with edge excitations.\refto{Hald}

Now the question is how to construct the Hilbert space of
these zero-energy edge states. One approach is to use the
current algebra techniques in CFT.  Let us concentrate
on the holomorphic part of the wave function ($\Phi_m$ or
$\t\Phi_m$ in \(1) and \(3)). To construct the edge states we need
to construct holomorphic functions that satisfies \(4).
We notice that the Laughlin-Jastrow function
$\Phi_m$ can be written as a correlation of the vertex
operators (or primary fields), $\psi_e$, in the
Gaussian model as follows:\refto{MR,Fubini}
$$\Phi_m=
\lim_{z_\infty\to \infty} (z_\infty)^{2h_N}
\<\Psi_N(z_\infty) \prod_{i=1}^N \psi_e(z_i)\>~,
\eqno(5)$$
where
$$ \psi_e(z)=e^{i\sqrt{m} \phi(z)},\ \ \ \ \
 \Psi_N(z)=e^{-iN\sqrt{m} \phi(z)}~.
\eqno(6)$$
Here $N$ is the number of electrons,
$h_N=mN^2/2$ the conformal dimension of $\Psi_N$,
and the scalar field, $\phi$, in the Gaussian model
is normalized so that $\< \phi(z)\phi(0) \>=\ln z$.
The factor $(z_\infty)^{2h_N}$ is included in \(5) so that the limit
$z_\infty\to \infty$ gives rise to a finite function.\refto{com}
Let $j(z)\equiv \partial \phi$ is the $U(1)$ current in
the Gaussian model. Since $j$ is
a local operator, we find that the correlation
$$\t\Phi_m\propto
\< \oint dz \a(z) j(z)
\Psi_N(z_\infty)\prod_{i=1}^N \psi_e(z_i)\>~,
\eqno(7)$$
with appropriately chosen holomorphic function $\a (z)$,
satisfies \(4) because of the OPE of $\psi_e(z)$:
$$\eqalign{
\psi_e(z)\psi_e(0)\propto & z^m e^{i2\sqrt{m} \phi(z)} +O(z^{m+1}) ~,\cr
j(z)\psi_e(0)\propto & {1\o z} \psi_e(0)+O(1) ~.\cr}
\eqno(8)$$
The integration contour of $z$ in \(7) encloses
all the $\psi_e(z_i)$ operators but not $\Psi_N$.
If $\a(z)$ has no poles except at $z_\infty$, then
$\(7)$ is finite for finite $z_i$, since $z_\infty\to \infty$.
Therefore \(7) is an eligible wave function for
the edge excitations. Introducing
$$j(z)=\sum_{n=-\infty}^\infty {j_n\o (z-z_\infty)^{n+1} }
\eqno(9)$$
and choosing
$$\a(z)=(z-z_\infty)^{-n} \ \ \ \ \ \ (n\geq 0)~,
\eqno(10)$$
we find that, by shrinking the contour around $z_\infty$
and letting $z_\infty\to \infty$,
\(7) becomes
$$\Phi_m^{(n)}=\lim_{z_\infty\to \infty} (z_\infty)^{2h_N+2n}
\<\Psi_N^{(n)} (z_\infty) \prod_{i=1}^N \psi_e(z_i)\>~,
\eqno(11)$$
where
$$\Psi_N^{(n)} (z_\infty)= j_n \Psi_N(z_\infty)~.
\eqno(12)$$
Again the factor $z_\infty^{2h_N+2n}$ is included in \(12)
to ensure the existence of a finite limit.\refto{com}
One can easily show that $j_n$ satisfy the $U(1)$ Kac-Moody
(KM) algebra
$$[j_n,j_m]=n\de_{m+n}~.
\eqno(12a)$$

The above discussion can be generalized to the case with
several insertions of the current
operator. The general statement is the following.
The ground state wave function can be written as a correlation
of $\psi_e(z_i)$ with a primary field $\Psi_N$ inserted
at infinity. To generate edge excitations above the ground
state, we simply replace the primary field $\Psi_N$
by its current descendants\refto{KZ},
which are generally of the form (with $n_1, n_2, \cdots, \geq 0$)
$$\Psi_N^{(n_1,n_2,...)} = (j_{-n_1}j_{-n_2}... )\Psi_N ~.
\eqno(13)$$
Let us call $l=\sum_i n_i$ the level of the descendant field
$\Psi_N^{(n_1,n_2,...)}$. The wave function
associated with it is given by
$$\Phi_m^{(n_1,n_2,..)}=\lim_{z_\infty\to \infty}
(z_\infty)^{2h_N+2l} \<\Psi_N^{(n_1,n_2,..)} (z_\infty)
\prod_{i=1}^N \psi_e(z_i)\>~.
\eqno(11a)$$
Such wave function has zero energy and can be identified
as an edge excitation.

We would like to show that all the
edge states generated by the level-$l$ descendant fields have
a total angular momentum $L=M_0+l$, where $M_0=mN(N-1)/2$ is the
total angular momentum of the ground state. We first note
that the angular momentum of the ground state
can be expressed in terms of conformal
dimensions, $h_e$ and $h_N$,
of $\psi_e$ and $\Psi_N$. Under a conformal transformation
$z\to w=f(z)$, the correlation of primary fields satisfies
$$\<\Psi_N(z_\infty) \prod_{i=1}^N \psi_e(z_i)\> =
(f'(z_\infty))^{h_N}\prod_i (f'(z_i))^{h_e}
\<\Psi_N(w_\infty) \prod_{i=1}^N \psi_e(w_i)\>~.
\eqno(13a)$$
Choosing $f(z)=\la z$, we have
$$\<\Psi_N(\la z_\infty) \prod_{i=1}^N \psi_e(\la z_i)\>
=\la^{-h_N-Nh_e} \<\Psi_N(z_\infty) \prod_{i=1}^N \psi_e(z_i)\>~,
\eqno(13b)$$
which implies that
$$\Phi_m(\la z_i)=\la^{h_N-Nh_e} \Phi_m(z_i)~.
\eqno(13c)$$
For $\la = e^{i\theta}$, this transformation is nothing
but a rotation by angle $\theta$ in the complex plane.
Thus the angular momentum of the ground state is
$$M_0=h_N-Nh_e~.
\eqno(13d)$$
With $h_N=mN^2/2$ and $h_e=m/2$, we see that $M_0=mN(N-1)/2$ as
expected for the Laughlin $1/m$-state. More generally,
since the dimension of a current descendant
$\Psi_N^{(n_1,n_2,...)}(z)$ is\refto{KZ} the sum $h_N+l$,
where $l$ is its level, we find that under $z\to w=\la z$
$$\Psi_N^{(n_1,n_2,...)}(z)=\la^{h_N+l}
\Psi_N^{(n_1,n_2,...)}(w)~;
\eqno(13e)$$
Thus, from \(11a) and \(13e) we can see
that the edge-excited state described by $\Phi_m^{(n_1,n_2,..)}$
carries an angular momentum of $L=h_N+l-Nh_e=M_0+l$.
As a general rule, valid for any FQH states that can be
generated by CFT, the angular momentum of an edge excitation
is equal to that of the ground state plus the level of the
descendant level of the associated insertion at infinity.

Note that the descendant fields $\Phi_m^{(n_1,n_2,..)}$
with different $(n_1,n_2,..)$ may not be linearly independent.
{}From the $U(1)$ KM algebra \(12a), it is easy to
show that the number, $D_l$, of linearly independent
descendant fields, \(13), at level $l$ is given by the
partition number of $l$. Mathematically this fact is
expressed in terms of the character (of the irreducible
representation, with the highest weight state $\Psi_N$)
of the $U(1)$ KM algebra as follows\refto{ch}:
$$\ch_N(\xi)\equiv \xi^{h_N} \sum_l D_l \xi^l
= \xi^{h_N} {1\over \prod_{n>0} (1-\xi^n)}\,.
\eqno(add1)
$$
If there is a one-to-one correspondence between the
edge states and the descendant fields of $\Psi_N$,
then we can use the above formula to obtain the number,
$D_L$, of edge excitations at any given angular
momentum $L$:
$$\eqalign{
\Ch_N(\xi) &\equiv \sum_L D_L \xi^L
= \ch_N (\xi) \xi^{-Nh_e} \cr
&= {1\over \prod_{n>0} (1-\xi^n)}\, \xi^{M_0}\,.}
\eqno(add2)
$$

The one-to-one correspondence between the edge states
and the descendant fields of $\Psi_N$ means that
different descendant fields always generate linearly
independent wave functions of $z_i$ through \(11a).
Since in \(11a), the descendant field
$\Psi_N^{(n_1,n_2,...)}(z_{\infty})$ acts on the state
$\prod_{i=1}^N \psi_e(z_i)$, we do not expect
this correspondence to be true for finite N and
arbitrarily large level $l=\sum_k n_k$. On the other hand,
it is conceivable that this one-to-one correspondence
holds for any finite level $l$ when $N$ is very large or
in the limit $N\to \infty$. Though we do not know,
at present, how to prove
this statement within the CFT approach for a
generic abelian FQH state, its validity for the
$1/m$-states can be seen in the following way:
It is known\refto{Hald,Stone} that the edge
states can be generated by multiplying
the ground state wave function by symmetric polynomials
of electron coordinates $z_i$. Mathematically,
the number of linearly independent symmetric
polynomials of degree $l$ is precisely given
by the partition number of the integer $l$.
Therefore, there is a close correspondence
between the descendant fields in the chiral
OPA in the bulk CFT description and the spectrum
of the edge states. For Laughlin $1/m$-states,
the bulk chiral OPA is simply a U(1) current
algebra\refto{MR,Fubini,Naples}; the correspondence
between the bulk CFT and the edge spectrum
has been known before\refto{Wp}.
In next section, we will show that this relationship
can be generalized to non-abelian FQH systems.

\head{3.  Edge excitations and operator product algebra}

In the last section we have generated the edge excitations of the
Laughlin state by inserting the current operator. One may try to
generate edge states by inserting the energy-momentum tensor $T$,
because the insertions of the energy-momentum tensor also maintain
the structure of zeros \(4) of the wave functions for the Laughlin
state. But one can show that the edge states
generated by the energy-momentum tensor is contained
in those generated by the current since $T\propto j^2$.
In the following we will discuss the edge excitations
of the Haldane-Rezayi (HR) state, in which case the
energy-momentum tensor does generate new edge states.

The HR state is a d-wave-paired spin-singlet FQH state
for spin-1/2 electrons. Apart from the usual Gaussian
factor, the holomorphic part of the wave function is given by
$$\eqalign{
\Phi_{HR}(z_i,w_i)=&\Phi_m(z_i,w_i)\Phi_{ds}(z_i,w_i)~, \cr
\Phi_{ds}(z_i,w_i)=& \cA_{z,w}\left( {1\o (z_1-w_1)^2}
{1\o (z_2-w_2)^2}...\right)~, \cr
\Phi_m(z_i,w_i)=& \left( \prod_{i<j} (z_i-z_j)^m
\prod_{i<j} (w_i-w_j)^m \prod_{i,j} (z_i-w_j)^m \right) \cr}
\eqno(a11)$$
which has a filling fraction $1/m$ with $m$ an even integer. Here
$z_i$ ($w_i$) are the coordinates of the spin-up
(spin-down) electrons, and $\cA_{z,w}$ is an operator
which performs separate antisymmetrizations
between $z_i$'s and between $w_i$'s. One can directly check
that $\Phi_{ds}$ is indeed a spin singlet and $\Phi_m$,
when viewed as an operator, commutes with the total spin operator.

Let us first analyze the structure of zeros of the HR wave
function assuming, for simplicity, $m=2$. Let
$z_1=z_2+\de_1$ and $z_1=w_1+\de_2$, we find
$\Phi_{HR}$ has the following expansion
$$\eqalign{
\Phi_{HR}=& \sum_{k=odd}(\de_1)^k A_{l}(z_2,..;w_1,..)~; \cr
\Phi_{HR}=& \sum_{n}(\de_2)^n B_{n}(z_2,..;w_1,..)~. \cr}
\eqno(h7b)$$
One can directly check that the coefficients
$$A_{1}=B_{1}=0~.
\eqno(h7c)$$
Therefore the HR state is the exact ground state of the following
two-body Hamiltonian
$$H=-V_1 \part_{z_1^*}    \de(z_1-z_2) \part_{z_1}
    -V_2 \part_{z_1^*}    \de(z_1-w_1) \part_{z_1}~,
\eqno(h7d)$$
since $H$ is positive definite for $V_i>0$, and
$\Phi_{HR}$ has a zero average energy. It has been checked
numerically that $\Phi_{HR}$ is the unique incompressible
ground state of $H$.\refto{HR} Other zero-energy states all
have higher angular momenta and are identified as
the edge excitations of the ground state.

It was shown in \Ref{WW} that the so-called
non-abelian part, $\Phi_{ds}$, can be written as a
correlation of two spin-$1/2$ primary fields $\psi_\pm$
in a $c=-2$ CFT. Therefore the ground state wave
function can be written as
$$\Phi_{HR}=\lim_{z_\infty\to \infty} (z_\infty)^{2h_N}
\< \Psi_N(z_\infty)
\prod_{i=1}^{N/2} \psi_{e+}(z_i)\psi_{e-}(w_i)\>~,
\eqno(14)$$
where
$$\psi_{e\pm}(z)=\psi_\pm (z) e^{i\sqrt{m} \phi(z)}~,
\eqno(h5)$$
$$\Psi_N(z_\infty)=e^{-iN\sqrt{m} \phi(z_\infty)}~.
\eqno(15)$$
Here $\phi(z)$ is the chiral scalar field in the Gaussian model
that  reproduces the $U(1)$ part $\Phi_m$. In \(14)
$h_N=mN^2/2$ is the dimension of $\Psi_N$ and $N$ the total number
of spin-up and spin-down electrons.
As shown in \Ref{WW}, the OPE of $\psi_e$ (for $m=2$)
$$\eqalign{
 \psi_{e+}(z+\de_1) \psi_{e+}(z) \propto &
 (\de_1)^3 \psi_1^1(z) e^{2i\sqrt{m} \phi(z)}+ O((\de_1)^4)~, \cr
 \psi_{e+}(z+\de_2) \psi_{e-}(z) \propto &
  e^{2i\sqrt{m} \phi(z)} + O((\de_2)^2)~, \cr}
\eqno(h7)$$
guarantees the condition \(h7c) due to the absence
of terms linear in $\de_1$ and $\de_2$.

Let $T$ be the energy-momentum tensor of the $c=-2$ model and
$j$ the $U(1)$ current of the Gaussian model.
The insertion of $T$ and $j$ does not affect the local OPE of
the $\psi_e$ operators, and hence the structure of local zeros
of the wave function is not affected.
Thus we can use both $T$- and $j$-insertions to generate
edge excitations that are zero-energy states of
the Hamiltonian \(h7d). Repeating the derivation in the
last section, we obtain edge excitations that are generated
from the descendant fields of $\Psi_N$:
$$\Phi_{HR}^{edge}=\lim_{z_\infty\to \infty}
(z_\infty)^{2h_N+2l} \< \Psi_N^{(n_1,..;m_1,..)}(z_\infty)
\prod_{i=1}^{N/2} \psi_{e+}(z_i)\psi_{e-}(w_i)\>~,
\eqno(16)$$
where
$$l=\sum_i n_i~+~\sum_k m_k
\eqno(16a)$$
is the level of the descendant field
$$\Psi_N^{(n_1,..;m_1,..)}(z_\infty)
=(L_{-n_1}...j_{-m_1}...)\Psi_N(z_\infty).
\eqno(17)$$
Here the fields $\Psi_N^{(n_1,..;m_1,..)}(z_\infty)$ are descendants
of $\Psi_N$ generated by the $U(1)$ current algebra of $j$ (formed by
$\phi$) and the $c=-2$ Virasoro algebra of $T$ (formed by
$\psi_{\pm}$ only). $L_n$ are the Fourier components of $T$:
$T(z)=\sum_{-\infty}^{+\infty} L_n/z^{n+2}$, which form
the $c=-2$ Virasoro algebra
$$[L_n,L_m]=(n-m)L_{n+m}+{c\o 12} n(n^2-1)\de_{m+n}~.
\eqno(18L)$$
Note both $T$ and $j$ are spin singlets. So they generate only
the spin-singlet sector of the edge excitations.

The space of the descendant fields is isomorphic to
the direct product of the representation of the KM
algebra generated by $j$ and the Verma module
of the identity operator generated by $T$.
This is because $T$ and $j$ commute with each other.
Similar to the pure U(1) case in the last section,
the edge excitations generated by level-$l$ descendant fields
can be shown to have angular momentum $L=M_0+l$, where
$M_0=h_N-Nh_e={m\o 2}N(N-1)-2N$ is the angular
momentum of the ground state $\Phi_{HR}$. Here we
have used \(13d), and $h_e={m\o 2}+1$,
$h_N={m\o 2}N^2$ (see \ref{WW}).

To study the spin-$s$ sector of the edge excitations,
we need to study the zero-energy states with total
spin $s$. The minimum angular momentum state with
total spin-$s$ and $S_z=\si$ ($s=1/2,1,3/2,...$) is given by
$$\Phi_{HR,s,\si}=\lim_{z_\infty\to \infty}
(z_\infty)^{2h_{N,s}} \< \Psi_{N,s,\si}(z_\infty)
\prod_{i=1}^{N_+}\psi_{e+}(z_i)\prod_{i=1}^{N_-}
\psi_{e-}(w_i)\>~.
\eqno(14a)$$
Here
$$\Psi_{N,s,\si}(z_\infty)=\psi_\si^{s}
e^{-iN\sqrt{m} \phi(z)}~,
\eqno(15a)$$
$h_{N,s}$ is the dimension of $\Psi_{N,s,\si}$, and
$N_+$ ($N_-$) is the number of spin-up (spin-down)
electrons. $N_\pm$ satisfy $N_+ -N_-=2\si$ and
$N$ is total numbers of the electrons $N=N_+ +N_-$.
The state \(14a) satisfies the condition \(h7c) and is
a zero-energy state. $\psi_\si^{s}$ is the spin-$s$
primary field discussed in \Ref{WW}, which is
generated from the multiple OPE of
$\psi_{\pm}\equiv\psi^{1/2}_{\pm 1/2}$
and has a conformal dimension
$$h_s={1\o 8} [(4s+1)^2-1].
\eqno(15b)$$
Thus $h_{N,s}=h_N+h_s={m\o 2}N^2+{1\o 8} [(4s+1)^2-1]$.
Other zero-energy states are generated by the
descendant fields of $\Psi_{N,s,\si}$:
$$\Psi_{N,s,\si}^{(n_1,..;m_1,..)} (z_\infty)
=(L_{-n_1}...j_{-m_1}...)\Psi_{N,s,\si}(z_\infty)~.
\eqno(17a)$$
The wave functions generated by them have the form
$$\Phi_{HR,s,\si}^{(n_1,..;m_1,..)}
=\lim_{z_\infty\to \infty} (z_\infty)^{2h_{N,s}+2l}
\< \Psi_{N,s,\si}^{(n_1,..;m_1,..)}(z_\infty)
\prod_{i=1}^{N_+}\psi_{e+}(z_i)\prod_{i=1}^{N_-}
\psi_{e-}(w_i)\>.
\eqno(14b)$$
They correspond to edge excitations with total
angular momentum
$$L=h_{N,s}+l-Nh_e ={m\o 2}N(N -1)+h_s-N+l~,
\eqno(18)$$
where $l$ is the descendant level given by \(16a).

To study the structure of the space of the descendant fields
$\Psi_{N,s,\si}^{(n_1,..;m_1,..)}$
generated from the primary field $\Psi_{N,s,\si}$, it is useful
to introduce the character associated to the primary
field $\Psi_{N,s,\si}$ and its descendants, defined by
$$\ch_{N,s}(\xi)=\sum_n D_n \xi^n~,
\eqno(19)$$
where $D_n$ is the number of the independent
descendant fields of scaling dimension
$n$. (Recall that the descendant field
$\Psi_{N,s,\si}^{(n_1,..;m_1,..)}$ has a scaling
dimension  $h_{N,s}+l$, and the character is independent
of the $S_z$ quantum number $\si$. Mathematically,
this is the character for the irreducible representation
of the $U(1)$ KM algebra and the Virasoro algebra with
the primary field $\Psi_{N,s,\si}$ as highest weight.)
Since $j$ acts only on $\Psi_N=e^{-iN\sqrt{m}\phi}$
and $T$ only on $\psi^s_\si$, the character
of $\Psi_{N,s,\si}=\Psi_N\psi_\si^s$ is the product of
the characters of the $\Psi_N$ and $\psi_\si^s$:
$$\ch_{N,s}(\xi)=\ch_N(\xi)~ \ch_s(\xi)~.
\eqno(20)$$
The character for the $U(1)$ KM algebra is the same as
\(add1)\refto{ch}:
$$\ch_N (\xi)={\xi^{mN^2/2}\o \prod_n (1-\xi^n)}~.
\eqno(chn)
$$
The character, $\ch_{s}$, in the $c=-2$ model,
is for the Virasoro representation
with the highest weight $[(4s+1)^2-1]/8$
(conformal dimension of $\psi^s_\si$).
By applying formula (19) given in \Ref{ch}, we obtain
$$\ch_s(\xi)={\xi^{h_s}-\xi^{h_{s+{1/2}}}
\o \prod_n (1-\xi^n)}~.
\eqno(21)$$
We notice that this character is non-trivial, because
of the existence of the null states.

Now let us summarize, in general terms but with the HR
state as an example, our above procedure for generating
edge excitations of a non-abelian FQH state
using OPA techniques.
{}For an FQH state that admits a OPA description\refto{WW},
in general, the electron operator(s) contain
two factors, the abelian part and non-abelian part
(see \(h5)). The abelian part is a vertex operator
in a Gaussian model, and the non-abelian part, e.g. for
the HR state, is $\psi_\pm$ which generate a $c=-2$ CFT.
As discussed in \Ref{WW}, $\psi_\pm$ in the non-abelian
part generate a closed OPA, called the center algebra,
through their OPE. The center algebra
contains the identity operator, $\psi_\pm$ and their
descendants, and all primary fields generated by
the OPE of $\psi_\pm$ (which in the HR case are the
spin-$s$ fields $\psi^s_{\si}$) and their descendants.
According to our chiral OPA description, the holomorphic
part of the FQH wave function (of zero energy)
can be written as a correlation between $\psi_{e\pm}$'s
with an operator $\Psi$ inserted at infinity.
The insertion $\Psi$ also contains two factors,
the abelian part
and a non-abelian part. The non-abelian part can be any
operator in the center algebra and the
abelian part can be any  descendent of
$\Psi_N$ under the $U(1)$ current $j$. If we choose
the non-abelian part of $\Psi$ to be
the identity operator and the abelian part to be
$\Psi_N=e^{-iN\sqrt{m} \phi(z)}$,
then $\Psi=\Psi_N$ will generate
the ground state wave function of the FQH state.
If the non-abelian part is chosen to be some other
operators in the center algebra and/or the
abelian part is chosen to
be an descendent of $\Psi_N$, then the insertion
$\Psi$ will generate the edge excitations.\refto{com1}

Having edge states generated in this way, two
questions immediately come to our mind: First,
are the edge states generated with different
descendants linearly independent to each other?
Second, do the insertions with all descendant
fields in the center algebra and the $U(1)$ current
algebra exhaust all
possible edge states? These are very hard questions.
Let us examine them in turns.

For the first question, obviously the descendants of
primary fields with different dimensions (which are
related to the angular momentum quantum numbers)
or different spin quantum numbers
generate linearly independent edge
states.
The hard part of the question is whether different
descendants of the same spin quantum number at the same level
will generate linearly independent states or not. Generally
in CFT, linearly independent descendants,
as operators, should have different (or linearly
independent) sets of correlations which contain {\it arbitrary}
numbers of electron operators $\psi_\pm$.
But when the number, $N$, of electrons is finite and
fixed, one can {\it not} claim that different
descendant fields generate
linearly independent correlations,
in particular for descendant fields at arbitrarily
large levels. So we suggest that the insertions
with different operators in the center algebra generate
different edge excitations in the thermodynamic limit or
in the large-$N$ limit; though at the moment we do
not know how to prove it within CFT. With the help
of the suggested correspondence between the descendant
fields and the edge states (in the large $N$ limit),
one can use the known results about the descendant
fields derived from the KM algebra and the center algebra
to obtain the
spectrum of low-lying edge states (see below).
We have done numerical diagonalization for small
systems to test the predictions. As will be seen in
Sec. 4, indeed the numerical results shows, on one hand,
the violation of the suggested correspondence
for finite $N$ at large level $l$. On the other hand,
they verify the validity of the correspondence
at the levels less than a certain number of order
$N$ in all FQH states we have considered.

Now we turn to the question of whether the descendant fields
discussed above can generate all possible edge excitations
in the system. We would like to first point out that the
edge wave functions constructed above not only preserve
the structure of zeros in the ground state as two electrons
approach each other, they may also preserve the structure
of higher-order zeros for three or more electrons
approaching each other. If the wave functions generated
by the descendant fields do not exhaust the zero-energy
sector of a two-body Hamiltonian, in principle it is possible
to construct a more restrictive Hamiltonian that contains,
in addition to the two-body interaction shown in \(h7d), also
three-body, four-body, \etc interactions, so that
the new Hamiltonian makes the wave functions
constructed above be and exhaust its zero-energy
edge-excited states. Thus, the question of whether
the space of edge states contains more states depends
on the dynamics of electron interactions, and cannot be
addressed by merely studying the wave functions.
In the following, we will assume that the Hamiltonian
satisfies the above conditions.

Under the assumption that the descendant fields and the edge states
have a one-to-one correspondence and the assumption that the
descendant fields exhaust all edge excitations,
the space of the edge excitations, $V_{edge}$,
can be written as
$$V_{edge}=V_{U(1)}\otimes V_{ca}
\eqno(22)$$
where $V_{U(1)}$ is the space of states of the $U(1)$ KM
algebra generated by $j_n$, and $V_{ca}$ the space of states
of the center algebra generated by $\psi_\pm$. Let us
introduce the character for edge excitations
$$\Ch(\xi)\equiv \sum_L D_L \xi^L~,
\eqno(23)$$
where $D_L$ is the number of the edge states with angular momentum
$L$. From \(18) and \(19), we see that the edge excitations of an
$N$-electron system with total spin $s$ and a fixed $S_z$ component
$\si$ are described by the following character
$$\Ch_{N,s}(\xi) =\ch_{N,s}(\xi)~ \xi^{-N h_e}
={1-\xi^{2s+1} \o \prod_n (1-\xi^n)^2}~
\xi^{M_0^{(s)}}~,
\eqno(24)$$
where $M_0^{(s)}=h_{N,s}-Nh_e$ is the minimum angular momentum
in the spin-$s$ sector. In section 4 we will test the
edge-state spectrum \(24) for the Hamiltonian \(h7d) by numerical
diagonalization.

The same approach can also be applied to the
p-wave\refto{MR,Wp,RR} and d-wave\refto{WW} paired FQH state
of spinless electrons.
Here we only present the final results.

Both pairing states contain a $Z_2$ structure,
because of a similar structure in their chiral OPA
(see \ref{WW}), corresponding to an
even or odd number of electrons. The center
algebra of (the non-abelian part of) the p-wave paired
state is generated by the dimension-1/2
(Majorana) fermion field in the Ising model,
which contains an even(odd)-sector generated by an
even (odd) number of fermion fields. (The fermion number
is not conserved, but is {\it mod 2} conserved.)
The character for the two sectors is given by that
for the $c=1/2$ Virasoro representation with the
identity ($h=0$) and the fermion field ($h=1/2$)
as the highest weight respectively. Using the formulas
(21-23) in \Ref{ch}, we obtain
$$\eqalign{
\ch_{even}=& {\sum_{m\in Z} \xi^{12m^2+m}
(1-\xi^{6m+1}) \o \prod_{n>0} (1-\xi^n)}~~~ ,\cr
ch_{odd}=& \xi^{1/2} {\sum_{m\in Z} \xi^{12m^2+5m}
(1-\xi^{6m+2}) \o \prod_{n>0} (1-\xi^n)}~~~ .\cr}
\eqno(25)$$
Thus the characters for edge excitations of the
p-wave paired state, after including the
Gaussian (or abelian) part, are
$$\eqalign{
\Ch_{e}=&{\sum_{m\in Z} \xi^{12m^2+m}
(1-\xi^{6m+1}) \o \prod_{n>0} (1-\xi^n)^2}~
\xi^{M_0^{(e)}}~~,
\ \ \ \hbox{for even number of electrons;} \cr
\Ch_{o}=& {\sum_{m\in Z} \xi^{12m^2+5m}
(1-\xi^{6m+2}) \o \prod_{n>0} (1-\xi^n)^2}~
\xi^{M_0^{(o)}}~~,
\ \ \ \hbox{for odd number of electrons.} \cr}
\eqno(26)$$
where $M_0^{e,o}$ is the angular momentum for the ground state
with an even or odd number of electrons.

As for the d-wave paired state, the center algebra
(for the non-abelian part)
is generated by a dimension-1
$U(1)$ current $\t j$ (which is different from the $U(1)$ current $j$
for the abelian part). In \Ref{WW} it was shown that the chiral
OPA of this $c=1$ model is a $Z_2$ orbifold model.
Each of the two sectors generated by even or
odd number of $\t j$'s may contain a lot of Virasoro
representations, since the OPE of two $\t j$'s may generate
higher integer-dimensional primary fields.
In the Appendix, we show that the characters
in the two sectors are given by
$$\ch_{even}={\sum_{m\geq 0} (-)^m\xi^{m^2}
\o \prod_{n>0} (1-\xi^n)},\ \ \ \
\ch_{odd}={\sum_{m\geq 1} (-)^{m-1}\xi^{m^2}
\o \prod_{n>0} (1-\xi^n)}~.
\eqno(27)$$
Note the sum of the above two characters is just the
character of the $U(1)$ KM algebra $1/\prod_{n>0} (1-\xi^n)$.
The chiral OPA of the $U(1)/Z_2$ model plus the abelian
$U(1)$ part leads to the following characters for edge
excitations of the d-wave paired state:
$$\eqalign{
\Ch_{e}(\xi)=&{\sum_{m\geq 0} (-)^m \xi^{m^2}
\o \prod_{n>0} (1-\xi^n)^2}~ \xi^{M_0^{(e)}}~,
\ \ \ \hbox{for even number of electrons}~; \cr
\Ch_{o}(\xi)=&{\sum_{m\geq 1} (-)^{m-1} \xi^{m^2-1}
\o \prod_{n>0} (1-\xi^n)^2}~ \xi^{M_0^{(o)}}~,
\ \ \ \hbox{for odd number of electrons}~. \cr}
\eqno(28)$$

Again in next section we will present a numerical test
of these predictions by directly diagonalizing the p-wave
and d-wave Hamiltonians given in \ref{Wp} and \ref{WW}.
As we will see, the numerical results for the
first a few low-lying edge states completely
agree with the CFT results \(24), \(26) and \(28). This
suggests that the chiral OPA algebra is an effective and
powerful tool to study edge excitations of
non-abelian FQH states.

\head{4. Numerical results and specific heat}

In this section we would like to compare the above chiral OPA
predictions with results of numerical diagonalization.
The edge excitations for the p-wave paired
FQH state have been studied in \ref{Wp}.
The edge spectrum of states with
low-lying angular momenta obtained numerically agrees
exactly with \(26) from chiral OPA.

For the HR state, according to \(24),
the number of edge excitations in the spin-$s$ sector
at angular momentum $L=M_0^s+l$,
with $M_0^s={m\o 2}N(N-1)+h_s-N$ (the minimum angular
momentum in the spin-$s$ sector) and $0\leq l\leq 5$,
is given by
$$\matrix{l:& 0&1&2&3& 4& 5 & \hbox{spin}  \cr
            & 1&1&3&5&10&16 & s=0\         \cr
            & 1&2&4&8&15&26 & s=1/2        \cr
            & 1&2&5&9&18&31 & s=1\         \cr
            & 1&2&5&10&19&34 & s=3/2        \cr
         }
\eqno(4.1)$$
We have
directly diagonalized of the Hamiltonian \(h7d) in the
symmetric gauge. The energy spectrum is labeled by the total
angular momentum. The energy eigenstates can be divided
into two classes. The first class is the zero-energy states
that starts at the angular momentum $M_0^s$.
Those states are identified as the edge excitations.
The second class is the bulk excitations with finite
energies. There is a clear energy gap that separates the edge
excitations and the bulk excitations, as an evidence for the
incompressibility of the HR state.
By counting the zero-energy states of $N$ electron system,
we find the following edge spectrum
$$\matrix{l:& 0&1&2&3&4    & \hbox{spin}  & \hbox{number}  \cr
            & 1&1&3&5&9    & s=0\         & N=6            \cr
            & 1&2&4&8&14   & s=1/2        & N=7            \cr
            & 1&2&5&8&14   & s=1\         & N=6            \cr
            & 1&2&5&9&     & s=3/2        & N=7            \cr
         }
\eqno(4.2)$$
The discrepancy between \(4.1) and \(4.2) for large $l$
is an effect due to finite N, as discussed in the previous
section. However, for $l\leq N_m\equiv (N-2s)/2$
the edge spectrum in the spin-$s$ sector has reached
the thermodynamical values (\ie those which do not change
when we increase $N$). From our numerical results for different
values of $N$, we also find that
the number of the edge states
at $l=N_m+1$ is always just one less than its
thermodynamical value. From \(4.2), we see that the thermodynamical
values of the edge spectrum, at least for low-lying angular
momentum states, well agree with the chiral OPA prediction \(4.1).

For the d-wave paired state for spinless electrons,
according to \(28), the number of edge excitations for an
even (or odd) number of electrons with low-lying angular
momenta $L=M_0^{(e)}+l$ (or $L=M_0^{(o)}+l$)
is given by
$$\matrix{l:& 0&1&2&3&4 & 5 & \hbox{sector} \cr
            & 1&1&3&5&11&18 & \hbox{even}   \cr
            & 1&2&5&9&18&31 & \hbox{odd}    \cr }
\eqno(4.3)$$
Direct diagonalizations of the d-wave Hamiltonian in \ref{WW}
for $N$ electrons give the following edge spectrum
$$\matrix{l:& 0&1&2&3& 4& 5  & \hbox{number}   \cr
            & 1&1&3&5&10&15  & N=6             \cr
            & 1&2&5&9&17&    & N=7             \cr
         }
\eqno(4.4)$$
For $l\leq N_m\equiv [N/2]$ the numbers of the edge states
have reached their thermodynamical value. The number
of the edge states at $l=N_m+1$ again is always just one less
than its thermodynamical value. We see again
that the thermodynamical values of the edge spectrum,
at least for low-lying angular momenta, well agree with
the results \(28) from chiral OPA.

Here we would like to remark that the
edge excitations in our calculations
have zero energy, because the our Hamiltonians do not contain
confining potential. If we include a parabolic confining
potential, then the Hamiltonian will receive a term
$V_c\hat L$ where $\hat L$ is the total angular momentum
operator. Thus the  parabolic confining
potential only shifts the energy without changing
the wave function. The edge excitations will
have a non-zero energy $V_c L=V_c(M_0 +l)$.
But the degeneracy at each total angular momentum remains
exactly the same. For more general confining potential
the degeneracy at a fixed angular momentum may be
lifted. Even in this case the edge excitations and the bulk
excitations remain well-separated. This is
because the energy $E_l$ of the edge
excitations of level $l$ is proportional to the
inverse of the length of the edge: $E_l\propto l/\sqrt{N}$.
When $l \ll \sqrt{N}$ the energy of the edge excitations is
much less than the bulk energy gap, and the edge and the bulk
excitations are still well-separated. In this case the chiral OPA,
instead of generating zero-energy states, generates
low-lying edge excitations.

Finally let us calculate the specific heat of
the edge excitations from the character formula.
Assume all edge excitations have the same
velocity $v$. The energy of edge excitations
in the angular momentum $L$ sector is given by
$$E=v(L-M_0)/R~,
\eqno(4.5)$$
where $M_o$ is the angular momentum of the ground state
and $R$ the radius of the FQH droplet.
Thus the partition function of edge excitations
is directly related to the character of the edge excitations
(see \(23)):
$$Z=\sum_{l=0} D_{l+M_0} e^{-l\be {v\o R}}
=\Ch(\xi)\xi^{-M_0}~,
\eqno(4.6)$$
with $\xi= e^{-\be {v\o R}}$. For one-dimensional systems,
the number of states, $N_n\equiv D_{n+M_0}$,
at level $n$, has the following asymptotic form
$$N_n\sim A n^\eta \exp({\sqrt{ {2\pi^2\o 3} c n}})~,
\eqno(4.7)$$
where $c$, $A$, and $\eta$ are constants.
{}From \(4.7) it follows that
the specific heat per unit length is
$$C=c{\pi\o 6} {T\o v}~,
\eqno(4.8)$$
independent of exponent $\eta$ of the prefactor in \(4.7).
The constant $c$ is known to be unity for the Gaussian
model\refto{Affl}. We will call $c$ the number of the
edge branches, since $c=n$ if edge excitations
are described by $n$ branches of chiral phonons (\ie
by $U(1)^n$ KM algebra).

The edge excitations for the d-wave FQH state contain a $U(1)$ part
and a non-abelian part. Let us concentrate on the non-abelian part.
First the specific heat of the non-abelian sector cannot be
larger than that of the Gaussian model:
$C\leq {\pi\o 6} {T\o v}$, since
the number of the edge excitations at each angular momentum
is less than that of the Gaussian model.
We also see from the character
formula \(27) that the number of the d-wave
edge excitations at level $n$ satisfies
$N^d_n> N^G_n-N^G_{n-1}$, where
$N_n^G\sim A_Gn^{\eta_G}\exp({\sqrt{ {2\pi^2\o 3} n} })$
is the number of the edge excitations
in the Gaussian model ($N^G_n$ is the number of partions
of the integer $n$).
Thus $N^d_n>A'n^{\eta_G-{1\o2}} \exp({\sqrt{ {2\pi^2\o 3} n} })$.
This implies $C\geq {\pi\o 6} {T\o v}$. So we have
$C={\pi\o 6} {T\o v}$. Including the $U(1)$ part,
the d-wave FQH state has a total of two branches
of edge excitations.

For the HR state, if we fix $S_z=0$, the
discussions in the last paragraph
also apply. We find the specific heat for non-abelian part of
the edge excitations to be $C= {\pi\o 6} {T\o v}$. Hence the HR
state also has two branches of edge excitations.
We note that in our present case, the number of
branches of edge excitations, $c=2$, is not the same as
the central charge ($-2+1$) of the bulk CFT.

\head{5. Discussions}

In this paper we have proposed that there is a close
connection between the spectrum of FQH edge excitations
and the descendent fields generated by $U(1)$ current
algebra and the center algebra (the chiral OPA generated by the
non-abelian part of the electron operator). This
connection allows us to apply CFT techniques
to construct wave functions for edge excitations,
and use algebraic methods to enumerate the edge
states. The predictions of the edge spectra
from the chiral OPA were confirmed by
numerical calculations for all FQH states studied.

Previously a similar connection between the (minimal)
chiral OPA and the bulk FQH wave function was studied
in detail in \Ref{WW}. Together with the connection
proposed in this paper between the (minimal) chiral
OPA and the FQH edge excitations, we see that
indeed chiral OPA provides us a unified description
of bulk wave function, quasiparticle excitations,
and edge excitations.
This indicates that the topological orders in the FQH
systems are characterized by chiral OPA.

To conclude, a few remarks are in order. First, the
$c=-2$ CFT in our discussion of the HR state is known to
contain negative-norm states, while all bulk and edge
excitations in the HR state, of course, have
positive norm. This indicates clearly that
the inner product between physical edge states
and that between the corresponding states in the CFT
are not the same. The chiral OPA in the CFT is used
only to generate the wave functions and spectrum of
edge excitations; nothing is implied by this for
the inner product of the physical edge states.
We also like to point out that our construction
of the edge states relies only on the chiral OPA.
One can, in principle, construct edge states and calculate
the edge characters directly from
the chiral OPA, as we did in the
appendix, without even mentioning the Virasoro algebra.
For the FQH states we studied in this paper, the chiral OPA
happen to form a representation of the Virasoro algebra.
This additional information allows us to use the well-known
character formula for the Virasoro algebra to calculate
the edge characters.

Near the completion of the present paper,
we learned that one can
also use a scalar fermion theory to construct edge
excitations of the HR state.\refto{Rnab} We would like
to thank N. Read for communicating his result prior to
publication.

We would like to thank Aspen Center of Physics
for hospitality, where this work was initiated.
YH would like to thank department of physics at MIT for hospitality
where part of the work was done.
XGW is supported by NSF grant No. DMR-91-14553
and YSW by NSF grant No. PHY-9309458.
XGW also would like to  thank
A.P. Sloan Foundation for support.

\head{Appendix}

Let us first consider the character of the descendent
fields of the identity generated by the $U(1)$
current $j$ (see \(add1)). Note the character for states
generated by $j_m$ and its power (with $m$ fixed) is given by
$\sum_{n=0} (\xi^m)^n$. Thus the total character has a form
$$\prod_{m=1}^\infty \sum_{n=0}
(\xi^m)^n={1\o \prod_{m>0}(1-\xi^m)}~.
\eqno(A1)$$
The character counts the number of states
generated by $j$ at each level.
In the phonon language, the level is proportional
to the total energy of phonons created by $j_{-m}$'s.
Thus the character counts the
degeneracy at each energy level.

To obtain the character of the descendent fields (or the character
of states) generated by an even or odd numbers of $j$,
we may consider
$$f(\xi, \eta)=\sum_{m=1}^\infty \sum_{n=0} (\eta \xi^m)^n
={1\o \prod_{m>0}(1-\eta \xi^m)}
\eqno(A2)$$
Here we use the power of $\eta$ to count the number of the
$j$-operators used to create a particular descendent
field (or a state). Thus the characters for the even
and odd sectors are given by
$$
\ch_{even}={1\o 2}[f(\xi,1)+f(\xi,-1)],\ \ \ \
\ch_{odd}={1\o 2}[f(\xi,1)-f(\xi,-1)]~.
\eqno(A3)$$
Using an identity
$${1\o 2}\prod_{n>0}{1-\xi^n \o 1+\xi^n} +{1\o 2}
=\sum_{n=0}^\infty (-)^n \xi^{n^2}
\eqno(A4)$$
we obtain \(27) from \(A3). The identity \(A4) can be obtained
from a $\th$-function identity
$\theta_4(0;2\tau)=[\eta(\tau)]^2 / \eta(2\tau)$, where
$\eta(\tau)$ is the Dedekind function. (See \Ref{th})

We know the OPE of $j$ generate many primary fields of the Virasoro
algebra. In fact those primary fields $\phi_n$
can be labeled by an integer $n=0,1,..$. The dimension of $\phi_n$
is $h_n=n^2$. $\phi_0$ is the identity and $\phi_1$
is the current operator.
Comparing the character formula of $\phi_n$
(see the theorem 5 in Sec. 3 of \Ref{ch}) with \(27), we find
that $\phi_n$ carries a $Z_2$-charge $(-)^n$.

\references

\refis{Halp}
The importance of edge states in the (integral)
quantum Hall effect was first pointed out by
B.I. Halperin, \prb 25, 2185, 1982.

\refis{Been}
C.W.J. Beenakker, \prl 64, 216, 1990.

\refis{MacD}
A.H. MacDanold, \prl 64, 220, 1990.

\refis{Wen0}
X.G. Wen, \prl 64, 2206, 1990; \prb 41, 12838, 1990;
D.H. Lee and X.G. Wen, \prl 66, 1675, 1991.

\refis{Hald}
F.D.M. Haldane, \journal Bull. Am. Phys. Soc.,
35, 254, 1990.

\refis{Stone}
M. Stone, \journal Ann. Phys., 207, 38, 1991;
\prb 42, 8399, 1990.

\refis{WenRev}
For a review, see X.G. Wen, \journal Int. J. Mod. Phys.,
B6, 1711, 1992.

\refis{Wen1}
X.G. Wen, \prb 40, 7387, 1989;
X.G. Wen and Q. Niu, \prb 41, 9377, 1990;
X.G. Wen, \journal Int. J. Mod. Phys., B4, 239, 1990.

\refis{prop}
X.G. Wen, in \ref{Wen0} and \ref{WenRev}.
See also Y.S. Wu, {\it ``Topological Aspects of
the Quantum Hall Effect''}, in {\it Physics, Geometry
and Topology}, Proceedings of 1989 Banff Summer School;
ed. H.C. Lee (Plenum Press, 1990).

\refis{WZ}
X.G. Wen and A. Zee, \prb 46, 2290, 1992.

\refis{Naples}
G. Christofano, G. Maiella, R. Musto and F. Nicodemi,
\pl B262, 88, 1991;
\journal Mod. Phy. Lett., A6, 1779, 1991;
{\it ibid} {\bf A6}, 2985 (1991); {\it ibid} {\bf A7}, 2583 (1992).

\refis{Cern}
A. Cappelli, V.G. Dunne, C.A. Trugenberger and G. Zemba,
\np 398B, 531, 1993; A. Cappelli, C.A. Trugenberger and G. Zemba,
preprint MPI-Ph/93-75, DFTT-65/93, October 1993. In our opinion,
the unitary irreducible representations of the $W_{1+\infty}$,
which coincide with those of multiple $U(1)$ current algebras,
provides a complete classification only for abelian FQH states.

\refis{MR}
G. Moore and N. Read, \np B360, 362, 1991.

\refis{Fubini}
S. Fubini, \journal Int. J. Mod. Phys., A5, 3553, 1990;
S. Fubini and C.A. L\"utken, \journal Mod. Phys. Lett., A6, 487, 1991.

\refis{Wen3}
X.G. Wen, \prl 66, 802, 1991.

\refis{WB}
B. Blok and X.G. Wen, \np B374, 615, 1992.

\refis{Wp}
X.G. Wen, \prl  70, 355, 1993.

\refis{WW}
X.G. Wen and Y.S. Wu, MIT and Utah preprint, Sept. 1993.
cond-mat/9310027

\refis{BPZ}
A.A. Belavin, A.M. Polyakov and A.B. Zamolochikov,
\np B241, 333, 1984.

\refis{book}
For readers not familiar with CFT, we recommend the reprint
book, ``{\it Conformal Invariance and Applications to Statistical
Mechanics}'', ed. C. Itzykson, H. Saleur and J.B. Zuber,
(World Scientific, 1988); and the review article
by P. Ginsparg, in Lectures at Les Houches
Summer School (1988), Vol. XLIX, ed. E. Brez\'{i}n and J.
Zinn-Just\'{i}n (North Holland, 1989).

\refis{KZ}
V.G. Knizhnik and A.B. Zamolochikov,
\np B247, 83, 1984.

\refis{HR}
F.D.M. Haldane and E.H. Rezayi, \prl 60, 956, 1988;
{\bf 60}, E1886 (1988).

\refis{ch}
A. Rocha-Caridi, {\it ``Vacuum Vector Representations
of the Virasoro Algebra''}, in {\it Vertex Operators in
Mathematics and Physics''}, MSRI Publications \# 3
(Springer, Heidelberg, 1984), p. 451.

\refis{RR}
N. Read and E.H. Rezayi, Yale and CSU preprint, May 1993.

\refis{Rnab} N. Read, private communication.

\refis{th}
See Eq. (A.1), (A.7) on p. 611 in
{\it Statistical Field Theory}, vol. 2, by
C. Itzykson and J.-M. Drouffe (Cambridge Univ. Press, 1989).

\refis{com} Consider an operator $\Psi$ of dimension
$h$. The correlation function $\<\Psi(z) ...\>$ (here ``..."
representing other operators) as $z\to \infty$ is
proportional to $\<\Psi(z)\Psi^\dag(0)\>(1+o(z^{-1}))$,
where $\Psi^\dag$ is the conjugate of $\Psi$.
Thus $\<\Psi(z) ...\> \prop z^{-2h}$ as $z\to \infty$.

\refis{com1} A similar but more general description of our construction
can be given directly in terms of the electron operators without separating
them into abelian and non-abelian parts.

\refis{Affl}
I. Affleck, \prl 56, 746, 1986.

\refis{psd} F.D.M. Haldane, \prl 51, 605, 1983.

\refis{TK} S.A. Trugman and S. Kivelson, \pr B26, 3682, 1985;
V.L. Pokrovsky and A.L. Talapov, \journal J. Phys. C, 18, L691, 1985.

\endreferences

\end